\documentclass[aps,10pt,superscriptaddress,nofootinbib,nobibnotes,amsmath,showkeys]{revtex4-2}

\pdfoutput=1

\usepackage{graphicx}
\usepackage{bm}

\newcommand{\h}{\overline{\mathrm{H}}}
\newcommand{\anti}[1]{\overline{\mathrm{#1}}}
\newcommand{\pos}{\mathrm{e}^{+}}

\begin{document}

\title{Limit on the Electric Charge of Antihydrogen}

\author{A. \surname{Capra}}
\email[Email address: ]{acapra@triumf.ca}
\affiliation{TRIUMF, 4004 Wesbrook Mall, Vancouver BC, Canada V6T 2A3}

\author{C. Amole} 
\altaffiliation[Present Address: ]{Dept. of Physics, Queen's University, Kingston, Ontario K7L 3N6, Canada}
\affiliation{Dept. of Physics and Astronomy, York University, Toronto ON, Canada M3J 1P3}

\author{M.D. Ashkezari}
\altaffiliation[Present Address: ]{Dept. of Earth, Atmospheric and Planetary  Sciences,  MIT,  Cambridge, MA 02139, USA}
\affiliation{Dept. of Physics, Simon Fraser University, Burnaby BC, Canada V5A 1S6}

\author{M. Baquero-Ruiz}
\altaffiliation[Present Address: ]{Centre de Recherches en Physique des Plasmas (CRPP), EPFL, CH-1015 Lausanne, Switzerland}
\affiliation{Dept. of Physics, University of California at Berkeley, Berkeley, California 94720-7300, USA}

\author{W. Bertsche}
\affiliation{School of Physics and Astronomy, University of Manchester, Manchester M13 9PL, UK}
\affiliation{The Cockcroft Institute, Daresbury Laboratory, Warrington WA4 4AD, UK}

\author{E. Butler}
\affiliation{Centre for Cold Matter, Imperial College, London SW7 2BW, UK}
\affiliation{Physics Dept., CERN, CH-1211 Geneva 23, Switzerland}

\author{C.L. Cesar}
\affiliation{Instituto de F\'{\i}sica, Universidade Federal do Rio de Janeiro, Rio de Janeiro 21941-972, Brazil}

\author{M. Charlton}
\author{S. Eriksson} 
\affiliation{Dept. of Physics, College of Science, Swansea University, Swansea SA2 8PP, UK}

\author{J. Fajans} 
\affiliation{Dept. of Physics, University of California at Berkeley, Berkeley, California 94720-7300, USA} 
\affiliation{Lawrence Berkeley National Laboratory, Berkeley, California 94720, USA}

\author{T. Friesen}
\affiliation{Dept. of Physics and Astronomy, Aarhus University, DK-8000 Aarhus C, Denmark}

\author{M.C. Fujiwara} 
\author{D.R. Gill} 
\affiliation{TRIUMF, 4004 Wesbrook Mall, Vancouver BC, Canada V6T 2A3}

\author{A. Gutierrez} 
\altaffiliation[Present Address: ]{Dept. of Medical Physics and Biomedical Engineering, University College London, London WC1E 6BT, UK}
\affiliation{Dept. of Physics and Astronomy, Univ. of British Columbia, Vancouver BC, Canada V6T 1Z1}

\author{J.S. Hangst} 
\affiliation{Dept. of Physics and Astronomy, Aarhus University, DK-8000 Aarhus C, Denmark}
\affiliation{Physics Dept., CERN, CH-1211 Geneva 23, Switzerland}

\author{W.N. Hardy}
\affiliation{Dept. of Physics and Astronomy, Univ. of British Columbia, Vancouver BC, Canada V6T 1Z1}
\affiliation{Canadian Institute of Advanced Research, Toronto ON, Canada M5G 1ZA}

\author{M.E. Hayden}
\affiliation{Dept. of Physics, Simon Fraser University, Burnaby BC, Canada V5A 1S6}

\author{C.A. Isaac} 
\affiliation{Dept. of Physics, College of Science, Swansea University, Swansea SA2 8PP, UK}

\author{S. Jonsell}
\affiliation{Dept. of Physics, Stockholm University, SE-10691 Stockholm, Sweden}

\author{L. Kurchaninov} 
\affiliation{TRIUMF, 4004 Wesbrook Mall, Vancouver BC, Canada V6T 2A3}

\author{A. Little}
\affiliation{Dept. of Physics, University of California at Berkeley, Berkeley, California 94720-7300, USA}

\author{J.T.K. McKenna} 
\affiliation{TRIUMF, 4004 Wesbrook Mall, Vancouver BC, Canada V6T 2A3}

\author{S. Menary}
\affiliation{Dept. of Physics and Astronomy, York University, Toronto ON, Canada M3J 1P3}

\author{S.C. Napoli}
\affiliation{Dept. of Physics, College of Science, Swansea University, Swansea SA2 8PP, UK}

\author{P. Nolan}
\affiliation{Dept. of Physics, University of Liverpool, Liverpool L69 7ZE, UK}

\author{K. Olchanski} 
\author{A. Olin} 
\affiliation{TRIUMF, 4004 Wesbrook Mall, Vancouver BC, Canada V6T 2A3}

\author{A. Povilus} 
\affiliation{Dept. of Physics, University of California at Berkeley, Berkeley, California 94720-7300, USA}

\author{P. Pusa} 
\affiliation{Dept. of Physics, University of Liverpool, Liverpool L69 7ZE, UK}

\author{F. Robicheaux}
\affiliation{Dept. of Physics, Purdue University, West Lafayette, Indiana 1947907, USA}

\author{E. Sarid} 
\affiliation{Dept. of Physics, NRCN-Nuclear Research Center Negev, Beer Sheva IL-84190, Israel}

\author{D.M. Silveira}
\affiliation{Instituto de F\'{\i}sica, Universidade Federal do Rio de Janeiro, Rio de Janeiro 21941-972, Brazil}

\author{C. So}
\altaffiliation[Present Address: ]{Dept. of Physics and Astronomy, University of Calgary, Calgary AB, Canada T2N 1N4}
\affiliation{Dept. of Physics, University of California at Berkeley, Berkeley, California 94720-7300, USA}

\author{T.D. Tharp} 
\affiliation{Dept. of Physics, Marquette University, 1250 W Wisconsin Ave, Milwaukee, WI 53233}

\author{R.I. Thompson} 
\affiliation{Dept. of Physics and Astronomy, University of Calgary, Calgary AB, Canada T2N 1N4}

\author{D.P. van der Werf} 
\affiliation{Dept. of Physics, College of Science, Swansea University, Swansea SA2 8PP, UK}

\author{Z. Vendeiro} 
\affiliation{Dept. of Physics, University of California at Berkeley, Berkeley, California 94720-7300, USA}

\author{J.S. Wurtele} 
\author{A.I. Zhmoginov}
\affiliation{Dept. of Physics, University of California at Berkeley, Berkeley, California 94720-7300, USA}
\affiliation{Lawrence Berkeley National Laboratory, Berkeley, California 94720, USA}

\author{A.E. Charman}
\affiliation{Dept. of Physics, University of California at Berkeley, Berkeley, California 94720-7300, USA}

\begin{abstract}
The ALPHA collaboration has successfully demonstrated the production and the confinement of cold antihydrogen, $\h$. An analysis of trapping data allowed a stringent limit to be placed on the electric charge of the simplest antiatom. Charge neutrality of matter is known to a very high precision, hence a neutrality limit of $\h$ provides a test of CPT invariance. The experimental technique is based on the measurement of the deflection of putatively charged $\h$ in an electric field. The tendency for trapped $\h$ atoms to be displaced by electrostatic fields is measured and compared to the results of a detailed simulation of $\h$ dynamics in the trap. An extensive survey of the systematic errors is performed, with particular attention to those due to the silicon vertex detector, which is the device used to determine the $\h$ annihilation position. The limit obtained on the charge of the $\h$ atom is \mbox{$ Q = (-1.3\pm1.8\pm0.4)\times10^{-8}$}, representing the first precision measurement with $\h$.
\end{abstract}

\maketitle

\section{Introduction}
\label{intro}
The study of cold antihydrogen is an experimental area is rapid development. The ALPHA collaboration demonstrated the confinement of cold $\h$ by means of a neutral atom trap \cite{an11ca1s} and the first spectroscopy experiment performed by inducing microwave transitions \cite{am12rqttaa}.

For $\h$ to be neutral, the electric charge of the positron, $\pos$, and antiproton, $\anti{p}$, must be exactly opposite. Measuring the $\h$ electric charge represents a CPT test, when a similar measurement is performed on hydrogen and the charge of its constituents is inferred. In addition, $\h$ charge neutrality is related to another important aspect of the Standard Model, namely the cancellation of the \textit{quantum anomaly}. It can be shown that the gauge invariance is guaranteed only if this anomalous contribution cancels out. The cancellation occurs by summing together all the relevant Feynman diagrams, as shown with great care in \cite{pe95itqft705-707}. This is equivalent to add up the colour and weak charges of the quarks and the leptons within a \textit{generation}. One of the consequences of this fundamental structure is that the sum of the electric charges of the valence quarks in the proton with the charge of the electron gives zero. The same reasoning can be extended from hydrogen to $\h$. Therefore, the $\h$ electric charge is intimately connected with the structure of the Standard Model and so it is fundamentally important that it be measured experimentally.

\section{Description of the Measurement}
\label{sec:HbarQmeas}
\begin{figure}[h!]
 \centering
 \includegraphics[scale=0.35]{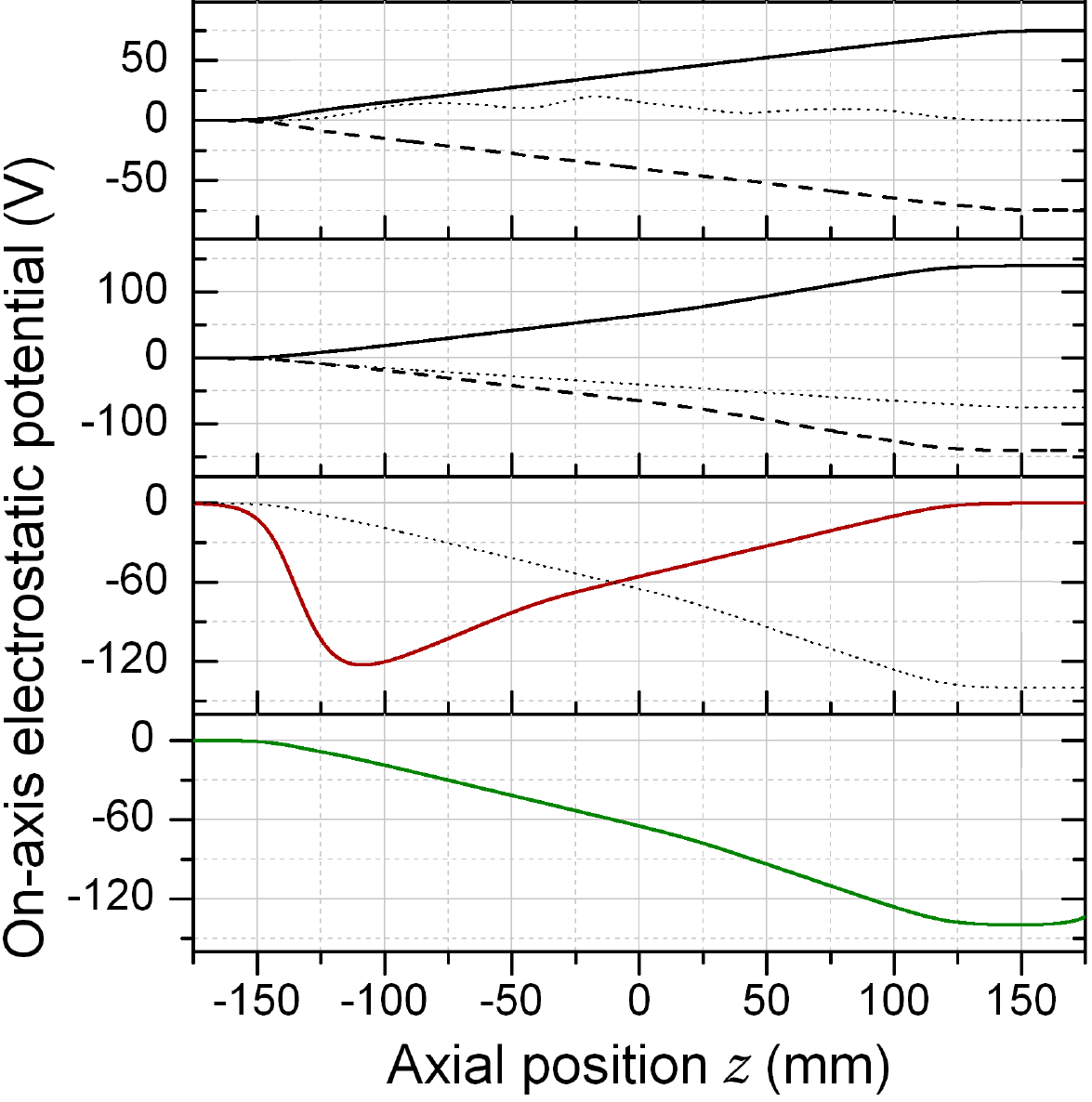}
 \caption[Clearing pulses and QWP.]{Black lines: clearing pulses of increasing strength to remove $\anti{p}$ from the mixing region. The dotted, dashed and solid lines indicate the potential at three different time instants. Red line: QWP \textit{bias-right}. Green line: QWP \textit{bias-left}. Adapted from \cite{ba13sna}.}
 \label{fig:biasV}
\end{figure}

Following $\h$ formation in the ALPHA trap \cite{am14aata}, a series of clearing electric pulses of increasing intensity is applied to remove $\anti{p}$ from the minimum-B trap (top two panels in Fig.~\ref{fig:biasV}). This procedure is adopted to ensure that no ``mirror-trapped'' $\anti{p}$ are left in the trap, as discussed in \cite{am12amaddbamamt}. Since ``bare'' or ``un-bound'' $\anti{p}$ are the major source of background, together with cosmics rays, to the detection of trapped $\h$, additional bias fields are applied when $\h$ is trapped in the magnetic trap and subsequently released. If any $\anti{p}$ remains in the trap, these bias fields, shown by the red and green lines in the bottom two panels of Fig.~\ref{fig:biasV}, cause them to drift towards one end of the trap in a given time, making them clearly identifiable against true $\h$ annihilation. Since $\h$ is released from confinement by ``quenching'' the superconducting magnets that produce the minimum-B trap, in the presence of the aforementioned bias fields, the technique is named \textit{quench with potential}, or QWP. 

The electric field in the QWP is called \textit{bias-right} $E_R$, if it sweeps the $\anti{p}$ to right-hand side of the trap\footnote{The trap axis coincides with the $z$-axis of a right-handed Cartesian coordinates system.}, i.e., \mbox{$z>0$}, and \textit{bias-left} $E_L$ to left-hand side of the trap, i.e., \mbox{$z<0$}.

If $\h$ has a charge $Qe$, where $e$ is the elementary charge and $Q$ is the \textit{fractional charge}, then the electric fields employed in the QWP would have a measurable effect on trapped $\h$. The present measurement is the search for a deflection of a trapped $\h$ in the $E_{L,R}$ fields due to its putative fractional charge
\begin{equation} \label{eq:HbarQrelLin}
 Q = s \left\langle z \right\rangle_{\Delta}\,,
\end{equation}
where $s$ is a parameter, dubbed \textit{sensitivity}, that depends only on constants and on the electric and magnetic fields present in the trap, and
\begin{equation} \label{eq:semidiff}
 \left\langle z \right\rangle_{\Delta} = \frac{\left\langle z \right\rangle_R - \left\langle z \right\rangle_L}{2}\,,
\end{equation}
is the deflection parameter, which has the advantage of cancelling out any offset, constant in time, of the averages \mbox{$\left\langle z \right\rangle_{L,R}$}, introduced in the measured $z$ of the antiatom annihilation position.

\section{Annihilation Data}
\label{sec:HbarQdata}
The $\h$ detection is accomplished by means of a position sensitive detector, the silicon vertex detector, or SVD, which determines the annihilation position, called the \textit{vertex} \cite{an12aarasd}. The average $z$ component, that is, the component along $E_{L,R}$, of $\h$ annihilation vertices, included in the present measurement is shown in Tab.~\ref{tab:dataHbarQ}.
\begin{table}[!h]
  \centering
  \begin{tabular}{|l|c|c|c|c|}
    \hline
    \textbf{Field Configuration}	&\textbf{Run 2010}	&\textbf{Run 2011}	&\textbf{Total} $\bm{\h}$	&$\bm{\left\langle z \right\rangle_{L,R}}$ \textbf{[cm]}\\
    \hline
    Bias Left				&145			&\_			&145				&$-0.02\pm0.53$\\
    Bias Right				&27			&214			&241				&$\phantom{-}0.79\pm0.42$\\
    \hline
  \end{tabular}
  \caption[Summary of detected $\h$.]{Summary of the $\h$ annihilation events, included in the present measurement.}
 \label{tab:dataHbarQ}
\end{table}

The value of Eq.~(\ref{eq:semidiff}), calculated from the experimental data, is
\begin{equation}\label{eq:shiftmeas}
 \left\langle z\right\rangle_{\Delta} = (0.4 \pm 0.3) \, \mathrm{cm}\,
\end{equation}
where the uncertainty is given by the propagation of the uncertainties on \mbox{$\left\langle z \right\rangle_{L,R}$}.

\section{Simulation and Sensitivity}
\label{sec:HbarQsim}
Since the electric and magnetic fields present in the trap are complex, the determination of the $Q$ requires numerical calculation of the $\h$ motion in the trap. This is performed with accurate modelling of all the known effects, such as electric field timing and intensity, magnetic field inhomogeneities, and SVD imperfections.

The relation between $Q$ and the deflection parameter, given by Eq.~(\ref{eq:semidiff}), is obtained from the numerical solution of the equation of motion of trapped $\h$, with the assumption that it has a non-zero electric charge $Qe$.

\begin{figure}[h!]
 \centering
 \includegraphics[scale=0.5]{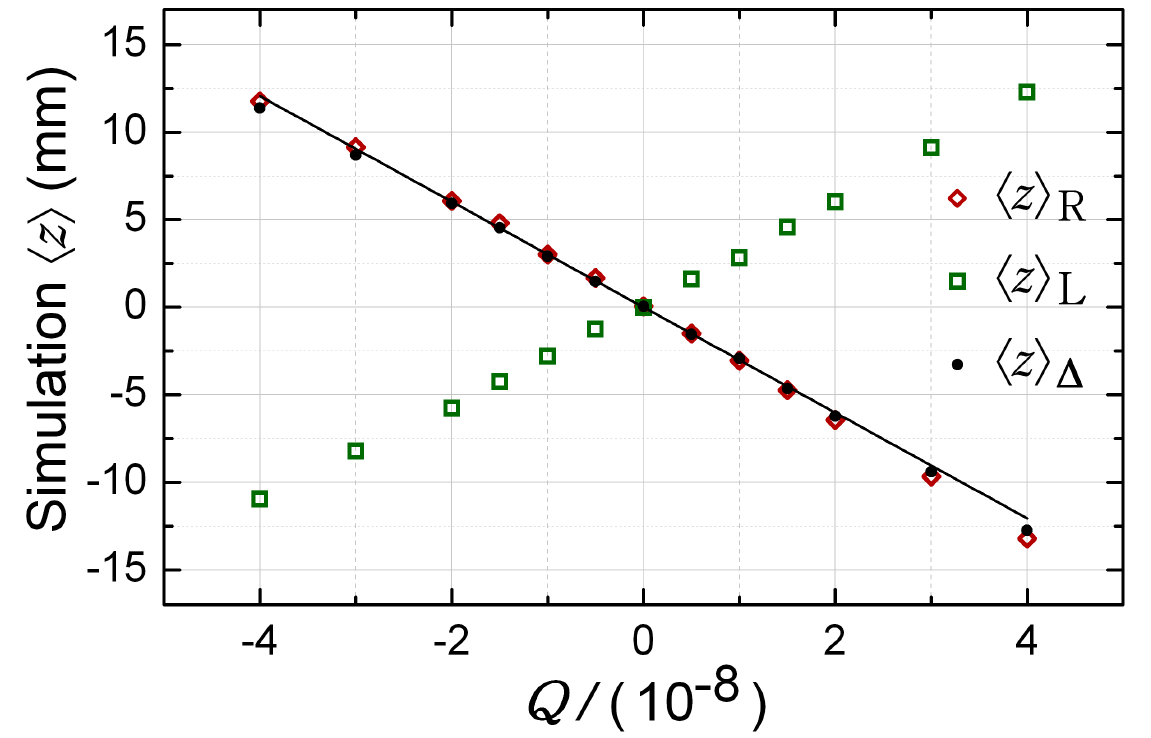}
 \caption[Relation between the $\h$ charge $Q$ and deflection parameter $\left\langle z \right\rangle_{\Delta}$.]{Results of the systematic study of effect of $Q$ (in units of $e$) on the deflection parameter $\left\langle z \right\rangle_{\Delta}$ (black dots). The resulting shift calculated with the two configurations of the bias voltages, \textit{left} and \textit{right}, are also shown as green boxes and red diamonds, respectively. The overlaid black line is the result of the best-fit. Taken from \cite{ba13sna}.}
 \label{fig:sens}
\end{figure}

The sensitivity, as defined by Eq.~(\ref{eq:HbarQrelLin}), is calculated from Fig.~\ref{fig:sens} by fitting a straight line to the \mbox{$(Q,\left\langle z \right\rangle_{\Delta})$} data and by taking the inverse of the slope:
\begin{equation}\label{eq:sens}
 s=\left( \frac{\mathrm{d}\left\langle z \right\rangle_{\Delta}}{\mathrm{d}Q}\right)^{-1} = (-3.31\pm0.04)\times10^{-9}\,\text{mm}^{-1} \,.
\end{equation}

Further details on the calculations discussed here can found in \cite{ba13sna}.

\section{Annihilation Vertex Detector Characterization}
\label{sec:SVDchar}
The accuracy of the determination the $z$ (axial) position of the annihilation vertex is very important, as it clearly impacts the measurement of the deflection $\left\langle z \right\rangle_{\Delta}$. Moreover, the dependece of the $\h$ detection efficiency on $z$ is also relevant, as it informs whether the measured deflection is due to SVD inefficiencies, rather than non-zero charge. A Monte Carlo simulation is used to estimate both the accuracy and the efficiency with which the $\h$ position is determined. These two quantities enter the calculation of the sensitivity $s$ and the systematic uncertainties.

The trapping experiments carried out with the bias-left configuration correspond roughly to the 2010 data collection run and the bias-right to the 2011 (as can be seen from Tab.~\ref{tab:dataHbarQ}). Hence, a change over time of the detection efficiency from one end of the detector with respect to the other may introduce a systematic shift in $\left\langle z \right\rangle_{\Delta}$, without necessarily implying non-zero $Q$. The evaluation of the SVD stability over the two years of operation crucially contributes to the calculation of the systematic uncertainties.

\begin{figure}[!h]
 \centering
  \includegraphics[scale=0.265]{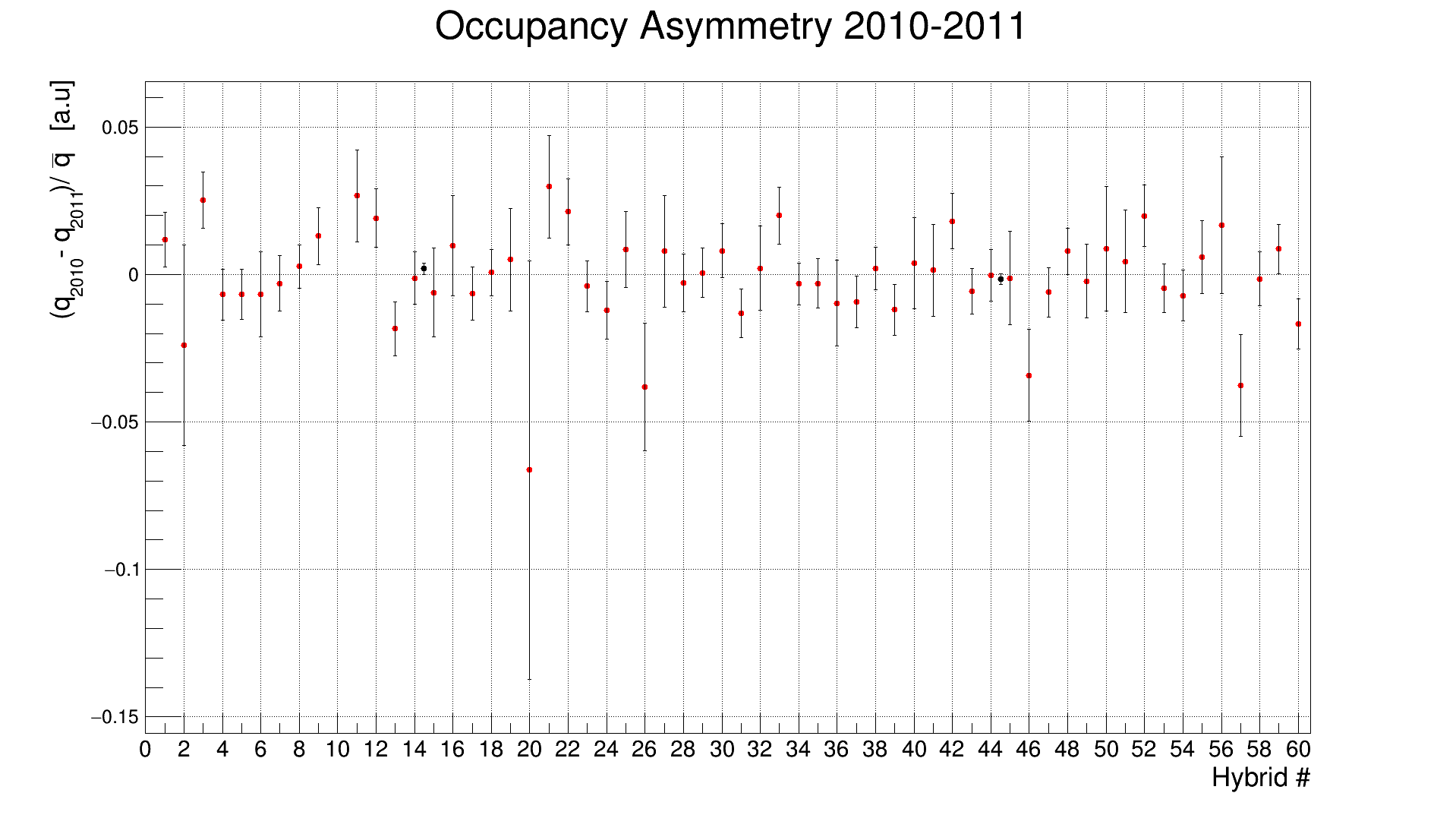}
 \caption[Occupancy change per hybrid.]{Red dots: occupancy change per hybrid given by Eq.~(\ref{eq:asymm}). Black Dots: occupancy asymmetry per detector half. Two points, corresponding to modules 2 and 20, seem to have an anomalous behaviour, but this is easily understood when it is recalled that module 10 was malfunctioning, i.e., disconnected. The latter, being in the middle layer between modules 2 and 20, reduces significantly the number of tracks that could pass across the other two modules, resulting in a built-in occupancy deficiency. Of course, this situation does not pose any problem at all, as long as it does not change over time.}
 \label{fig:asymm}
\end{figure}

The present analysis exploits cosmic rays in order to establish the single hybrid\footnote{A hybrid constitutes a single readout unit, or module, of the SVD.} \textit{occupancy}, that is, the number of \textit{hits per hybrid} in each data collection run, normalized to the total number of hits occurring in that particular run. An asymmetry parameter per hybrid per year is computed:
\begin{equation} \label{eq:asymm}
 a_h = \frac{q_{2010,h} - q_{2011,h}}{(q_{2010,h} + q_{2011,h})/2}\,,
\end{equation}
where $h=1,\ldots,60$ labels the hybrids and $q_{i,h}$ for the years $i=2010,2011$ is the occupancy level of the hybrid $h$. In other words, the difference in occupancy level between the two years divided by the mean occupancy level is used to asses the stability of a detector module, i.e., whether it shows an anomalous or asymmetric behaviour in time. The results are displayed in Fig.~\ref{fig:asymm}.

The average change in occupancy (black dots in Fig.~\ref{fig:asymm}) for the $z<0$ half is $(0.2 \pm 0.2)\%$ and for the $z>0$ half is $(-0.2 \pm 0.2)\%$. Both are consistent with no change in occupancy over time at the $0.2\%$ level and compatible with each other at $0.4\%$. A full analysis of the SVD is presented in \cite{capra15}.

\section{Result and Discussion}
\label{sec:HbarQres}
A number of sources can contribute to a drift over time of $\left\langle z \right\rangle_{\Delta}$ (e.g., the subject of the previous section) and to a change in $s$ with respect to its nominal value, given by Eq.~(\ref{eq:sens}) (e.g., an inaccurate knowledge of the current in the mirror coils). These effects are taken into account in the calculation of the systematic error of $Q$. The investigation of the various errors proceeded by comparing simulations with and without the inclusion of the proposed sources. 

By using the measured deflection of Eq.~(\ref{eq:shiftmeas}) and the simulation result of Eq.~(\ref{eq:sens}), the fractional charge of $\h$ from Eq.~\ref{eq:HbarQrelLin} is $Q=-1.3\times 10^{-8}$, which translates into the following limit
\begin{equation}\label{eq:HbarQlimit}
 |Q| < (1.3 \pm 1.1 \pm 0.4) \times 10^{-8}\,,
\end{equation}
where the first error is the statistical error due essentially to $\left\langle z \right\rangle_{\Delta}$, since the uncertainty on $s$ has a negligible contribution, and the second is the overall systematic error on $Q$. For a detailed discussion of the systematic uncertainties see \cite{am14elca}.

This limit is a million-fold improvement over the previous experimental result obtained with an energetic $\h$ beam \cite{gr97a}. The best limit on the neutrality of ordinary atoms and molecules, such as $\mathrm{He}$, $\mathrm{H}_2$ and $\mathrm{SF}_6$, is $10^{-21}e$, due, of course, to the availability of large quantities of those species \cite{br11tnmbamsr}, while less than 1000 antiatoms have been trapped to date (and then only singly). The accuracy on $Q$, Eq.~(\ref{eq:HbarQlimit}), is limited by statistics and is the first high precision measurement on antimatter. This result also represents a test of CPT invariance, provided that the hydrogen (fractional) charge is known.

The ALPHA collaboration has recently improved the limit on the $\h$ charge by means of a novel technique called \textit{stochastic acceleration}, that employs a series of randomly oscillating electric fields to ``walk'' the antiatoms out of the trap, if they are charged. This result is reported in \cite{ahmadi16}.

\bibliographystyle{plain}       

\end{document}